\documentclass[journal]{IEEEtran}
\usepackage{amsmath,amsfonts}
\usepackage{color,xcolor}
\usepackage{algorithm}
\usepackage{array}
\usepackage[caption=false,font=normalsize]{subfig}
\usepackage{textcomp}
\usepackage{stfloats}
\usepackage{url}
\usepackage{verbatim}
\usepackage{graphicx}
\usepackage{cite}
\usepackage{amsthm} 
\usepackage{mathtools}
\usepackage{amssymb}
\usepackage{algpseudocode}
\usepackage{booktabs}
\usepackage{multirow}
\usepackage{microtype}
\usepackage{hyperref}
\usepackage{enumerate}

\theoremstyle{definition}
\newtheorem{theorem}{Theorem}
\newtheorem{assumption}{Assumption}

\newtheorem{remark}{Remark}

\newtheorem{problem}{Problem}

\newcommand{\ie}{\textit{i.e.}}
\newcommand{\eg}{\textit{e.g.}}
\newcommand{\RR}{\mathbb{R}}
\newcommand{\CC}{\mathbb{C}}
\newcommand{\RH}{\mathcal{RH}}

\DeclareMathOperator{\diag}{diag}

\begin{document}

\bstctlcite{IEEEtranBSTCTL_no_dash}

\title{Modular Model Order Reduction for Inverter-Based Power Systems with Stability and Accuracy Guarantees}

\author{Hanqing Zhang,~\IEEEmembership{Graduate Student Member,~IEEE,}
Pudong Ge,~\IEEEmembership{Member,~IEEE,}
Mohammad Fahim Shakib,~\IEEEmembership{Member,~IEEE,}
Adri\`a Junyent-Ferr\'e,~\IEEEmembership{Senior Member,~IEEE,}
Balarko Chaudhuri,~\IEEEmembership{Fellow,~IEEE,}
Luuk Poort,
Lars A. L. Janssen,
Giordano Scarciotti,~\IEEEmembership{{Senior Member},~IEEE}
}

\markboth{Journal of \LaTeX\ Class Files,~Vol.~14, No.~8, August~2021}%
{Shell \MakeLowercase{\textit{et al.}}: A Sample Article Using IEEEtran.cls for IEEE Journals}


\maketitle

\begin{abstract}
This paper proposes a modular model order reduction framework for large-scale power grids integrated with renewable energy sources.
The proposed framework has the advantage of reducing individual sub-areas independently while retaining the power grid interconnection structure.
Moreover, given a user-defined accuracy specification on the overall reduced-order interconnected power grid, corresponding target specifications for the individual reduced sub-areas are determined: satisfaction of such individual specifications guarantees the overall user-defined accuracy specification,
while also preserving the stability of the overall reduced-order interconnected power grid.
The proposed framework is validated on a modified IEEE $118$-bus system (where synchronous generators have been replaced by inverter-based resources) by assessing the accuracy specification satisfaction of the overall reduced-order interconnected power grid, in terms of frequency response, stability, active power, and sub-synchronous behaviour when injected with disturbances.
\end{abstract}


\section{Introduction}\label{sec:introduction}

\IEEEPARstart{S}{imulation} and analysis of modern power systems require high-fidelity mathematical models to capture the intricate behaviour of power grids~\cite{simualtion_requirement}. The inherent complexity of large-scale interconnected power grids is continuously escalating due to the rapid integration of inverter-based resources (IBRs)~\cite{GPST_RAG,ps_high_ibr}.
This introduces three novel challenges relevant to complexity: (1) Scale: to replace synchronous generators, one needs hundreds or thousands of IBRs; (2) Timescale: fast-acting control loops blur the boundary between timescales; and (3) Richness: IBRs give rise to new phenomena such as sub-synchronous oscillations~\cite{Gu_ps_stability}.
Consequently, the introduction of IBRs induces an extreme computational burden when simulating and analysing the dynamical behaviour of the power grid for security assessment or real-time operation \cite{ps_dynamics}. To reduce this computational bottleneck, model order reduction (MOR) techniques, such as balanced truncation~\cite{freitas2008gramian,ali2018model} and moment matching~\cite{scarciotti2016low,gong2024model}, have been widely adopted to simplify the complexity of power grid models.

\begin{figure}[!t]
\centering
\subfloat[$\mathcal{G}$]{\includegraphics[width=0.5\columnwidth]{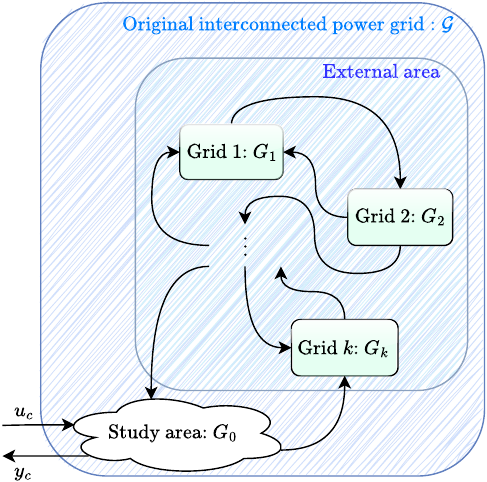}%
\label{fig:PowerGridFOM}}
\hfil
\subfloat[$\hat{\mathcal{G}}$]{\includegraphics[width=0.5\columnwidth]{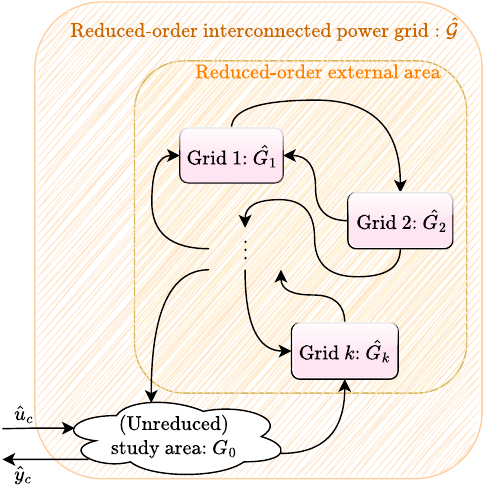}%
\label{fig:PowerGridROM}}
\caption{Schematic layout of (a) the original interconnected power grid, partitioned into a study area and an external area with the latter comprising multiple regional sub-grids; and (b) the reduced-order interconnected power grid assembled by reduced-order external-area sub-grids and the unreduced study area.}
\label{fig:PowerGridFOMROM}
\end{figure}

In the context of MOR for power systems~\cite{chow2013power},
a large-scale power grid is partitioned into a study area, whose mathematical description is maintained in full detail, and an external area, which consists of the remaining part of the network and is reduced to facilitate efficient simulation~\cite{gong2024model,leung2019model}.
The schematic procedure is illustrated in Fig.~\ref{fig:PowerGridFOMROM}.
Available MOR methods for power systems, see \eg~\cite{chaniotis2005model,sturk2014coherency,ali2018model,ali2020model}, usually treat the external area as one lumped region, thereby neglecting its multi-area structure.
However, in modern practice, the external area geographically spans multiple interconnected subsystems, such as individual regional grids managed independently, as shown in Fig.~\ref{fig:PowerGridFOMROM}.
Although structure-preserving techniques have been applied in~\cite{leung2019model,sturk2014coherency}, the computational resources needed for performing the MOR process scale up in line with the order of the overall interconnected system (rather than that of individual subsystems), which may become less computationally tractable for larger-scale power grids.
Consequently, as the power grid is inherently multi-area, reducing the entire interconnected system in one go remains computationally demanding. 
This necessitates a flexible, modular MOR framework that allows individual, independent, and thus computationally cheap reduction for regional sub-grids prior to system-wide integration.

In addition to modularity, another open problem in MOR for power systems is how to derive rigorous accuracy specifications and preserve stability of the overall reduced-order interconnected system (unreduced study area interconnected with reduced-order external area).
Although MOR techniques such as balanced truncation~\cite{ali2018model} can provide quantitative error bounds for the reduced-order external area and preserve stability, \textit{these guarantees are only valid when viewing the external area in isolation}.
Crucially, to the best of the authors' knowledge, no MOR method for power systems guarantees that the \textit{overall} reduced-order interconnected system (unreduced study area interconnected with reduced-order external area) remains stable and satisfies a user-defined accuracy specification.
This deficiency has so far hindered providing a rigorous accuracy analysis regarding the discrepancy between simulations of the original and the reduced-order interconnected power grids.

To address the aforementioned challenges, this paper introduces a modular MOR framework for large-scale interconnected power grids.
This framework adapts the method proposed in~\cite{janssen2024modular} and~\cite{janssen2023modular} to power systems by extending the method to a dynamical interconnection structure that includes the detailed description of the study area.
Note that the proposed method can be seen as a higher-level wrapper that can be used in combination with any standard MOR method applied to individual sub-grids. We demonstrate this characteristic by blending the Loewner framework with this modular approach, enabling the modular MOR framework to be applicable even when only the black-box frequency scans of the sub-grids are available, which may be the case due to IBRs.

The main contributions of this paper are summarised as follows.

\begin{itemize}
    \item We propose a modular MOR framework that preserves the physical boundaries of individual regional sub-grids. By preserving the physical interconnection structure between the sub-grids and retaining the exact physical meaning of tie-line and boundary input-output states, the generated reduced-order interconnected power grid model remains physically interpretable and directly applicable for further system analysis.
    \item We introduce a novel mathematical formulation that quantifies the reduction error propagation from the regional sub-grids to the overall interconnected power grid and \emph{vice versa}. This translates a user-defined, frequency-weighted global reduction accuracy specification for the overall interconnected power grid into actionable, specific local reduction accuracy specifications for individual external-area sub-grids.
    \item We provide an analytical guarantee that, upon reconnecting the unreduced study area to the independently reduced sub-grid models that satisfy their individual accuracy specifications, the reduced-order interconnected power grid (i) is guaranteed to be stable and (ii) satisfies the predefined overall accuracy specification.
    This ensures reliable and highly accurate dynamic simulations for the entire power grid.
    \item We develop a novel data-driven MOR method by combining the Loewner framework and the greedy algorithm to obtain reduced-order sub-grids that satisfy their local accuracy specifications.
\end{itemize}

Finally, we demonstrate the applicability and advantages of the proposed modular MOR framework on a modified IEEE $118$-bus benchmark model, where we replaced synchronous generators with grid-following (GFL) and grid-forming (GFM) inverters. Both frequency- and time-domain simulations are presented to validate a substantial reduction of the benchmark model while preserving modularity, stability, and prediction accuracy. 

The remainder of this paper is organised as follows. Section~\ref{sec:preliminaryAndProblemStatement} introduces some preliminaries and formulates the modular MOR problem. Section~\ref{sec:Framework} provides the modular MOR framework that guarantees specified reduction accuracy and stability. The simulation results are presented in Section~\ref{sec:case-study}, while Section~\ref{sec:conclusion} concludes the paper.

\textbf{Notation.} The symbols $\RR$ and $\CC$ denote the sets of real and complex numbers, respectively.
The imaginary unit is denoted by $\iota = \sqrt{-1}$, and for a complex number $z$, $|z|$ denotes its absolute value.
The identity and zero matrices are denoted by $I$ and $\bf{0}$, respectively, whose dimensions can be inferred from the context or subscript.
The real rational subspace of $\mathcal{H}_\infty$ is denoted by $\RH_\infty$, containing all proper and real rational stable transfer functions.
Given a transfer function {$G: \CC \to \CC^{p \times m} \in \RH_\infty$ with $s \in \CC$} the Laplace variable, $\|G(s)\|_{\mathcal{H}_\infty}$ denotes its $\mathcal{H}_\infty$-norm.
For a matrix $A \in \CC^{p \times m}$, $A^\star$ denotes its conjugate transpose and $\|A\|_2 = \bar \sigma(A)$ denotes its induced 2-norm with $\bar \sigma(A)$ the maximum singular value of $A$.

\IEEEpubidadjcol

\section{Preliminaries and Problem Statement}\label{sec:preliminaryAndProblemStatement}

In this section, we start by introducing the modular model of the power grid and reframe it as an upper linear fractional transformation (LFT) in Section~\ref{sec:subModularModelFramework}.
The derivation of the reduction accuracy specification is presented in Section~\ref{sec:subErrSpecificationRepresentation}.
Finally, the problem statement is formalised in Section~\ref{sec:subProblemStatement}.



\subsection{Modular Model Description}\label{sec:subModularModelFramework}

The power grid is constructed by multiple sub-grids that are managed individually with different operational boundaries, as shown in Fig.~\ref{fig:PowerGridFOM}.
Each sub-grid includes power generation (\ie, inverter-based resources), transmission lines, and power loads, whose dynamical behaviours are governed by associated nonlinear vector fields.
The interconnection of these sub-grid models results in a model of the overall interconnected power grid and its governing nonlinear vector field.




To analyse the oscillatory behaviour of the power grid around an operating mode, its model is linearised at the investigated operating point.
Thus, the power grid model is represented as an interconnection of $k+1$ linear time-invariant (LTI) subsystems with transfer functions $G_j: \CC \to  \CC^{p_j \times m_j}$, for $j \in \{0, \dots, k\}$.
The transfer function of the overall interconnected power grid model is $\mathcal{G}: \CC \to  \CC^{p_c \times m_c}$.
The input and output of each subsystem $j \in \{0, \dots, k\}$ are denoted by $u_j \in \RR^{m_j}$ and $y_j \in \RR^{p_j}$, respectively. 
The external input and output of the interconnected power grid model are defined as nodal disturbances $u_c \in \RR^{m_c}$ and states of interest $y_c \in \RR^{p_c}$, respectively.
We denote the study area as $G_0$ and the external area that comprises the remaining subsystems as $G_{j}$ for $j \in \{1, \dots, k\}$, as illustrated in Fig.~\ref{fig:PowerGridFOM}.

\begin{figure}[!t]
\centering
\subfloat[$\mathcal{G}$]{\includegraphics[width=0.8\columnwidth]{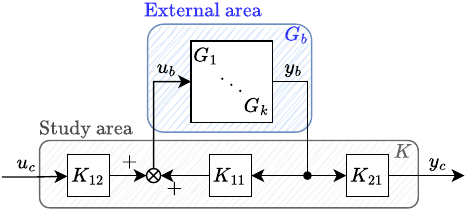}%
\label{fig:FOMLFT}}
\hfil
\subfloat[$\hat{\mathcal{G}}$]{\includegraphics[width=0.8\columnwidth]{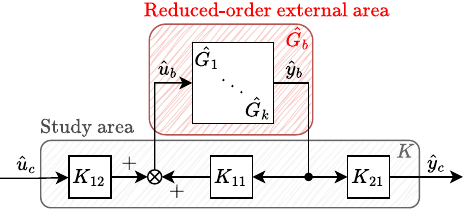}%
\label{fig:ROMLFT}}
\caption{Linear fractional transformation configurations of (a) the full-order model and (b) the reduced-order model, respectively, of the overall interconnected power grid.}
\label{fig:FOMROMLFT}
\end{figure}

To develop the modular MOR framework, we need to aggregate all interconnection relationships between subsystems into an interface model, allowing the power grid to be formulated as an upper LFT of the interface model and the collection of decoupled subsystems, see~\cite{janssen2024modular}.
However, unlike~\cite{janssen2024modular}, in which the interface model is static, we incorporate the study area into the interface model since it is unchanged (unreduced), resulting in a dynamic interface.
To this end, we collect the transfer functions of the external-area subsystems in block-diagonal (decoupled) form
\begin{equation}\label{eq:Gb}
    G_b \coloneqq \diag\left(G_1, \dots, G_k\right): \CC \to \CC^{p_b \times m_b}
\end{equation}
with $m_b \coloneqq \sum_{j = 1}^k m_j$ and $p_b \coloneqq \sum_{j = 1}^{k} p_j$, and, accordingly, the decoupled external-area model $G_b$ has the input $u_b \coloneqq \begin{bmatrix}
    u_1^\top, \dots, u_k^\top
\end{bmatrix}^\top \in \RR^{m_b}$ and the output $y_b \coloneqq \begin{bmatrix}
    y_1^\top, \dots, y_k^\top
\end{bmatrix}^\top \in \RR^{p_b}$.
The interconnection relationships between the external-area subsystems can then be extracted and fully characterised by the dynamic interface (containing the study-area dynamics) $K: \CC \to \CC^{(m_b + p_c) \times (p_b + m_c)}$ described by {(assuming no feedforward term from the external input $u_c$ to the external output $y_c$ without loss of generality)}
\begin{equation}\label{eq:interfaceK}
    K(s) = \begin{bmatrix}
        K_{11}(s) & K_{12}(s) \\
        K_{21}(s) & \bf{0}
    \end{bmatrix}, 
\end{equation}
with inputs $y_b$ and $u_c$ and outputs $u_b$ and $y_c$, see Fig.~\ref{fig:FOMLFT}.

To construct an explicit representation of $K$, assume that $(A_0, B_0, C_0)$ is a minimal state-space realisation of the study area, so that $G_0(s) = C_0 (sI -A_0)^{-1} B_0$.
Then, the state-space realisation $(A_K, B_K, C_K, D_K)$ of $K$ is given by
\begin{equation}\label{eq:determineInterfaceK}
    A_K = A_0, \quad B_K = B_0 P_B, \quad C_K = P_C C_0,
\end{equation}
where $P_B \in \RR^{m_0 \times (p_b + m_c)}$ and $P_C \in \RR^{(m_b + p_c) \times p_0}$ are permutation matrices that first augment the input and output space of the study area to $\RR^{(p_b + m_c)}$ and $\RR^{(m_b + p_c)}$, respectively, and then reorder the rows of $B_0$ and the columns of $C_0$ to align the places where the signals $u_0$ and $y_0$ appear in $[y_b^\top, u_c^\top]^\top$ and $[u_b^\top, y_c^\top]^\top$, respectively.
The matrix $D_K$ is a feedforward term.
Specifically, the $(k,j)$-element of $D_K$ is either $1$ or $0$, depending on whether the $j$-th entry of $[y_b^\top, u_c^\top]^\top$ is directly connected to the $k$-th entry of $[u_b^\top, y_c^\top]^\top$.
As a consequence, the transfer function of the dynamic interface is obtained as $K(s) = C_K (sI - A_K)^{-1} B_K + D_K$.

Now, the transfer function of the full-order model (FOM) of the overall interconnected power grid, from the external input $u_c$ to the external output $y_c$, can be equivalently reformulated as the upper LFT of $G_b$ and $K$, that is\footnote{See~\cite{zhou1998essentials} for a detailed description of linear fractional transformations.},
\begin{equation}\label{eq:Gc}
    \mathcal{G}(s) \coloneqq K_{21}(s) G_b(s) \left(I - K_{11}(s) G_b(s)\right)^{-1} K_{12}(s).
\end{equation}

We then seek to obtain a reduced-order model (ROM) of the overall interconnected power grid $\mathcal{G}$ as in~\eqref{eq:Gc} such that the physical interconnection structure between the subsystems is preserved, stability is retained, and user-defined accuracy specifications are satisfied.
Additionally, the MOR process is expected to be computationally tractable and applicable in a modular way, without the need to share full model information between the subsystems (sub-grids).



To enable the modular MOR feature, we reduce each subsystem independently and then reassemble the ROMs to obtain the reduced-order external area and thus the reduced-order interconnected power grid, while preserving the interconnection relationships between the subsystems.
In particular, each external-area subsystem $G_j$ for $j \in \{1, \dots, k\}$ is reduced to $\hat G_j: \CC \to \CC^{p_j \times m_j}$ with the same input-output channels
as $G_j$, \ie, $\hat u_j \in \RR^{m_j}$ and $\hat y_j \in \RR^{p_j}$, but of considerably smaller order.
Analogously, we collect these ROMs $\hat G_j$ in block-diagonal (decoupled) form and define
\begin{equation}\label{eq:hatGb}
    \hat G_b \coloneqq \diag (\hat G_1, \dots, \hat G_k): \CC \to \CC^{p_b \times m_b}
\end{equation}
with the input $\hat u_b \coloneqq \begin{bmatrix}
    \hat u_1^\top, \dots, \hat u_k^\top
\end{bmatrix}^\top \in \RR^{m_b}$ and the output $\hat y_b \coloneqq \begin{bmatrix}
    \hat y_1^\top, \dots, \hat y_k^\top
\end{bmatrix}^\top \in \RR^{p_b}$.
Since the interconnection structure between the external-area subsystems and the study area, \ie, the dynamic interface $K$, is preserved, the resulting ROM of the overall interconnected power grid is obtained by replacing each external-area subsystem $G_j$ with its ROM $\hat G_j$, as illustrated in Fig.~\ref{fig:PowerGridROM}.
Consequently, the transfer function of the reduced-order interconnected power grid model $\hat{\mathcal{G}}: \CC \to \CC^{p_c \times m_c}$ can be represented by the upper LFT of $\hat G_b$ and the same (dynamic) interface $K$ given by~\eqref{eq:interfaceK}-\eqref{eq:determineInterfaceK}, as shown in Fig.~\ref{fig:ROMLFT}, that is,
\begin{equation}\label{eq:hatGc}
    \hat{\mathcal{G}}(s) \coloneqq K_{21}(s) \hat G_b(s) (I - K_{11}(s) \hat G_b(s))^{-1} K_{12}(s).
\end{equation}

To ensure satisfaction of user-defined reduction accuracy specifications and stability, in the next section, we recall the reduction accuracy analysis techniques introduced in~\cite{janssen2023modular,janssen2024modular}.

\subsection{Unified Representation of Reduction Accuracy Specifications}\label{sec:subErrSpecificationRepresentation}

In this section, we introduce the mathematical formulation of the reduction accuracy specification for ROMs.
This formulation is subsequently used to establish the relationship between the reduction performance at a subsystem level and at an overall interconnected-system level.


We denote the reduction error dynamics of the overall interconnected power grid as
\begin{equation}\label{eq:Ec}
    E_c(s) \coloneqq \hat{\mathcal{G}}(s) - \mathcal{G}(s).
\end{equation}
Then, based on standard robust control tools, see~\cite{zhou1998essentials},
a (user-defined) accuracy specification for the ROM $\hat{\mathcal{G}}$ can be characterised by the set
\begin{equation}\label{eq:errSpecInterconnectedSys}
    \mathcal{E}_c(s) \coloneqq \{\hat{\mathcal{G}}(s) \mid \|W_c(s) (\hat{\mathcal{G}}(s) - \mathcal{G}(s)) V_c(s)\|_{\mathcal{H}_\infty} \leq 1\},
\end{equation}
where $W_c: \CC \to \CC^{p_c \times p_c}$ and $V_c: \CC \to \CC^{m_c \times m_c}$ are stable bi-proper minimal-phase weighting transfer functions.
This accuracy specification flattens and unifies the allowed reduction errors by pulling out scaling factors from $\hat{\mathcal{G}} - \mathcal{G}$ to $W_c$ and $V_c$.
From a power system perspective, such a frequency-weighted configuration has the benefit of providing flexibility in defining different reduction accuracy requirements (error bounds) for individual input-output channels in different frequency ranges.
In this way, the weighting transfer functions $W_c$ and $V_c$ can be used to make sure that the behaviour of the resulting ROM is close to the FOM within a range of frequencies.
The extent to which the ROM is close to the FOM and the frequency range can be arbitrarily enforced by selecting appropriate $W_c$ and $V_c$.
For instance, this may be useful to make sure that the ROM can reproduce critical dynamics with high fidelity, such as inter-area oscillations or sub-synchronous oscillations introduced by renewable generation (\eg, modes below $50$ Hz), by imposing stricter reduction accuracy requirements within frequency ranges associated with those phenomena.
Conversely, this allows one to relax accuracy constraints at frequencies irrelevant to the grid's electromechanical dynamics, thereby leaving more room for reduction.
How to select the weighting transfer functions $W_c$ and $V_c$ to enforce this is illustrated in Section~\ref{sec:subModularMORSimulation}.

Similarly, for each external-area subsystem $G_j$, where $j \in \{1, \dots, k\}$, we denote
\begin{equation}
    E_j(s) \coloneqq \hat G_j(s) - G_j(s)
\end{equation}
as its reduction error dynamics. Then the reduction accuracy specification for the ROM $\hat G_j$ can be represented by the set
\begin{equation}\label{eq:errSpecSubSys}
    \mathcal{E}_j(s) \coloneqq \{\hat{G}_j (s) \mid \|W_j(s) (\hat{G}_j(s) - G_j(s)) V_j(s)\|_{\mathcal{H}_\infty} < 1\},
\end{equation}
where $W_j:\CC \to \CC^{p_j \times p_j}$ and $V_j: \CC \to \CC^{m_j \times m_j}$ are stable bi-proper minimal-phase weighting transfer functions.
While it may not be clear at this point, the accuracy specifications~\eqref{eq:errSpecInterconnectedSys} and~\eqref{eq:errSpecSubSys} have clear graphical interpretations, as illustrated in the simulations reported later in this paper.



\subsection{Problem Statement}\label{sec:subProblemStatement}

In this section, we formulate the modular MOR problem that needs to be solved in this paper.
To streamline the presentation, we first pose the following assumption.

\begin{assumption}
    The overall interconnected power grid model $\mathcal{G}$ as in~\eqref{eq:Gc} is well-posed
    and internally stable~\cite[Chapter~$10$]{zhou1998essentials}.
\end{assumption}

This assumption ensures that the interconnected power grid model is computationally tractable (well-posed) and that all internal states remain bounded under perturbations injected from any interconnection port (internally stable).

We next formulate the problem to be addressed in this paper.

\begin{problem}\label{prob:I}
    Given a user-defined accuracy specification $\mathcal{E}_c$ for the reduced-order interconnected power grid, find reduction accuracy specifications $\mathcal{E}_j$ for the reduced-order subsystems $j \in \{1, \dots, k\}$ such that if $\hat G_j \in \mathcal{E}_j$ for all $j \in \{1, \dots, k\}$ then the resulting $\hat{\mathcal{G}} \in \mathcal{E}_c$, \ie{} the overall interconnected reduced-order power grid satisfies the predefined accuracy specification and preserves (internal) stability.
\end{problem}

Problem~\ref{prob:I} aims to determine the accuracy specification required for the subsystem $j \in \{1, \dots, k\}$ such that if all the ROMs $\hat G_j$ satisfy the determined accuracy specifications, the resulting reduced-order interconnected system $\hat{\mathcal{G}}$, constructed by integrating $\hat G_j$ with the unreduced study area, satisfies the user-defined accuracy specification $\mathcal{E}_c$ for the overall interconnection.
Next, we want to reduce all the individual external-area subsystems such that their respective accuracy specifications are satisfied. The associated problem is formulated as follows.

\begin{problem}\label{prob:III}
    Given the reduction accuracy specification $\mathcal{E}_j$ for the subsystem $j \in \{1, \dots, k\}$,
    construct the ROM $\hat G_j$ that satisfies the accuracy specification.
\end{problem}

In summary, the objective is to independently reduce the external-area subsystems while preserving the interconnection structure such that when the reduced subsystems are reassembled with the unreduced study area following the preserved interconnection structure, the resulting reduced-order interconnected power grid satisfies the predefined accuracy specification and preserves stability.

\section{Modular Model Order Reduction Framework}\label{sec:Framework}

By leveraging the upper LFT reformulation of the power grid model and the reduction accuracy specification established in Section~\ref{sec:preliminaryAndProblemStatement}, this section presents the modular MOR framework that constructively solves Problems~\ref{prob:I} and~\ref{prob:III}.
Firstly, a relationship between the accuracy specification of the overall interconnected power grid and those of the external-area subsystems is detailed in Section~\ref{sec:subErrorSpecificationTransformation}.
This relationship is later used to determine the admissible reduction accuracy specifications for the external-area subsystems.
Then, several MOR methods are proposed in Section~\ref{sec:subDecentralisedMOR4Subsystems} to determine the ROMs of the external-area subsystems such that the accuracy specifications are satisfied.
Finally, in Section~\ref{sec:subOverviewModularMOR}, we provide an overview of the modular MOR framework.

\subsection{Top-Down Accuracy Specifications Translation}\label{sec:subErrorSpecificationTransformation}

\begin{figure}[!t]
\centering
\includegraphics[width=0.85\columnwidth]{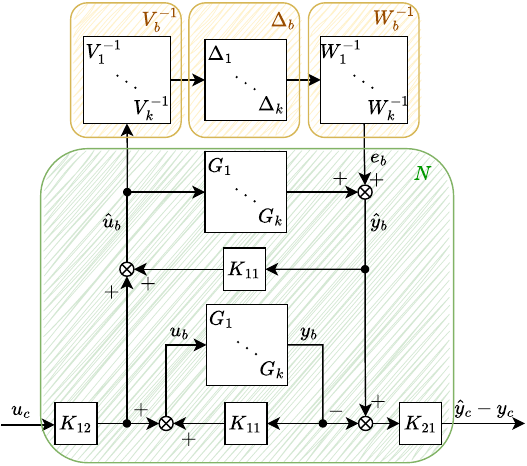}
\caption{Block diagram of the reduction error dynamics $\hat{\mathcal{G}} - \mathcal{G}$ of the overall interconnected power grid, represented as an upper LFT of the nominal system $N$ and the subsystem accuracy specifications $W_b$, $\Delta_b$, and $V_b$.\label{fig:addUncertaintyEc}}
\end{figure}

Without loss of generality, any element that belongs to the accuracy specification set~\eqref{eq:errSpecSubSys} enjoys the form~\cite{zhou1998essentials}
\begin{equation*}
    \hat G_j(s) = G_j(s) + W_j^{-1}(s) \Delta_j(s) V_j^{-1}(s),
\end{equation*}
with $W_j^{-1}: \CC \to \CC^{p_j \times p_j}$ and $V_j^{-1}: \CC \to \CC^{m_j \times m_j}$ the (stable proper) inverses of $W_j$ and $V_j$, respectively, and $\Delta_j: \CC \to \CC^{p_j \times m_j}$ any stable transfer function such that $\|\Delta_j\|_{\mathcal{H}_\infty} < 1$.
Thus, the external-area ROM collection $\hat G_b$ in~\eqref{eq:hatGb} can be rewritten as
\begin{equation}\label{eq:hatGbAddUncertainty}
    \begin{aligned}
        \hat G_b(s) 
        = G_b(s) + W_b^{-1}(s) \Delta_b(s) V_b^{-1}(s),
    \end{aligned}
\end{equation}
with
\begin{equation}
    \begin{aligned}
        &W_b^{-1} \coloneqq \diag (W_1^{-1}, \dots, W_k^{-1}): \CC \to \CC^{p_b \times p_b}, \\
        &\Delta_b \coloneqq \diag (\Delta_1, \dots, \Delta_k): \CC \to \CC^{p_b \times m_b}, \\
        &V_b^{-1} \coloneqq \diag (V_1^{-1}, \dots, V_k^{-1}): \CC \to \CC^{m_b \times m_b}.
    \end{aligned}
\end{equation}
By using~\eqref{eq:hatGbAddUncertainty}, the difference between $\mathcal{G}$ and $\hat{\mathcal{G}}$ can be expressed by the block diagram in Fig.~\ref{fig:addUncertaintyEc}.
In this diagram, we separate the reduction accuracy specifications of the external-area subsystems, namely $V_b^{-1}$, $\Delta_b$, and $W_b^{-1}$, from the remaining known dynamics, that is, the original external-area subsystem collection $G_b$ and the dynamic interface $K$ (comprising the unreduced study-area dynamics).
In this way, we obtain a propagation system in the green block that describes how the reduction accuracy specifications of the external-area ROMs propagate through the interconnected power grid and contribute to the overall reduction accuracy specification of the reduced-order interconnected power grid.
Following robust control conventions, we refer to the propagation system as the nominal system since it has fixed and known dynamics isolated from reduction accuracy specifications to be determined.
Let the nominal system be defined by $N: \CC\to \CC^{(p_c + m_b) \times (m_c + p_b)}$ as
\begin{equation}\label{eq:nominalSysN}
    N(s) \coloneqq \begin{bmatrix}
        N_{11}(s) & N_{12}(s) \\
        N_{21}(s) & \bf{0}
    \end{bmatrix},
\end{equation}
with inputs $u_c$ and $e_b$, where $e_b \in \RR^{p_b}$ is the output signal of the yellow reduction accuracy specification block $W_b^{-1} \Delta_b V_b^{-1}$, and outputs $\hat y_c - y_c$ and $\hat u_b$, see Fig.~\ref{fig:addUncertaintyEc}.
Comparing~\eqref{eq:nominalSysN} with the input and output flow in Fig.~\ref{fig:addUncertaintyEc} yields
\begin{equation}
    \begin{aligned}
        N_{11}(s) = (I - K_{11}(s) G_b(s))^{-1} K_{11}(s),\\
        N_{12}(s) = (I - K_{11}(s) G_b(s))^{-1} K_{12}(s),\\
        N_{21}(s) = K_{21}(s) (I - G_b(s) K_{11}(s))^{-1}.
    \end{aligned}
\end{equation}
Consequently, the reduction error $\hat{\mathcal{G}} - \mathcal{G}$ can be equivalently represented as an upper LFT of $N$ and $W_b^{-1} \Delta_b V_b^{-1}$,
where $N$ is a fully known transfer function, consisting of the dynamic interface $K$ and the external-area subsystem collection $G_b$.


\begin{figure}[!t]
\centering
\includegraphics[width=0.75\columnwidth]{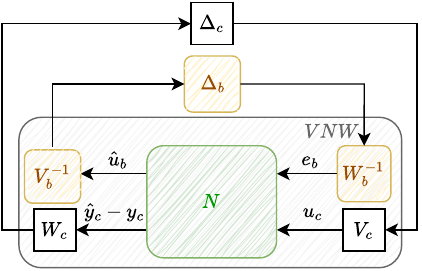}
\caption{Block diagram of the reduction error dynamics $\hat{\mathcal{G}} - \mathcal{G}$ of the overall interconnected power grid, weighted by the defined $W$ and $V$ and augmented by the fictitious block $\Delta_c$.\label{fig:mainLoop}}
\end{figure}

To relate the local accuracy specifications to the global one, we consider the block diagram in Fig.~\ref{fig:mainLoop}, where $N$ is weighted by $V$ and $W$ defined as
\begin{equation}\label{eq:VWWeighting4NominalSys}
    \begin{aligned}
        V &\coloneqq \diag(V_b^{-1}, W_c): \CC \to \CC^{(p_c+m_b) \times (p_c+m_b)}, \\
        W &\coloneqq \diag(W_b^{-1}, V_c): \CC \to \CC^{(m_c+p_b) \times (m_c+p_b)},
    \end{aligned}
\end{equation}
and $\Delta_c: \CC \to \CC^{m_c \times p_c} \in \RH_\infty$ is an augmented fictitious block such that $\|\Delta_c\|_{\mathcal{H}_\infty} \leq 1$.
Then, the user-defined accuracy specification $W_c$ and $V_c$ for the reduced-order interconnected power grid $\hat{\mathcal{G}}$ can be related to the subsystem accuracy specifications $W_b^{-1}$ and $V_b^{-1}$ through $N$ by the following theorem, which is adapted from~\cite[Thm.~$1$]{janssen2023modular}.

\begin{theorem}\label{thm:VWMainLoopThm}
    Let
    \begin{equation*}\label{eq:DMatrix4Optimisation}
    \begin{aligned}
        \mathcal{D}(\omega) &:= \Big\{(D_l(\omega), D_r(\omega)) \;\Big|\; d_1, \dots, d_k, d_c: \RR \to \RR_{>0}, \\
        &D_l(\omega) = \diag(d_1(\omega)I_{p_1}, \dots, d_k(\omega)I_{p_k}, d_c(\omega)I_{m_c}), \\
        &D_r(\omega) = \diag(d_1(\omega)I_{m_1}, \dots, d_k(\omega)I_{m_k}, d_c(\omega)I_{p_c})
        \Big\}.
    \end{aligned}
    \end{equation*}
    Given global accuracy specifications $W_c$ and $V_c$, if there exist $(D_l(\omega), D_r(\omega)) \in \mathcal{D}(\omega)$ and diagonal $W$ and $V$ defined in~\eqref{eq:VWWeighting4NominalSys} with positive real-valued entries such that
    \begin{equation}\label{eq:LMI4Optimisation}
        \begin{bmatrix}
            W^2(\iota \omega) D_r^{-1}(\omega) & N^\star(\iota \omega)\\
            N(\iota \omega) & V^2(\iota \omega) D_l(\omega)
        \end{bmatrix} \succ \bf{0}
    \end{equation}
    for all $\omega \in \RR$, then the property $\hat G_j \in \mathcal{E}_j$ for all $j \in \{1, \dots, k\}$ ensures the resulting $\hat{\mathcal{G}} \in \mathcal{E}_c$ and preserves (internal) stability.
\end{theorem}

This theorem relates subsystem and interconnected system accuracy specifications and presents a sufficient condition under which Problem~\ref{prob:I} is solved.
In practice, we check~\eqref{eq:LMI4Optimisation} on a fine grid of $\omega \in \Omega \subset \RR$ using the D-K iteration in~\cite{zhou1998essentials}.

\begin{remark}
    The relationship between $\mathcal{E}_c$ and $\mathcal{E}_j$ revealed in Theorem~\ref{thm:VWMainLoopThm} can also be used in the so-called bottom-up problem. 
    By solving that problem, one can determine a reduction accuracy specification for the overall reduced-order interconnected power grid based on the (known) reduction accuracy specifications of the external-area ROMs.
    Readers are referred to~\cite{janssen2024modular} for details.
\end{remark}

\subsection{Model Order Reduction Based on Accuracy Specifications}\label{sec:subDecentralisedMOR4Subsystems}

Now that we have a way to determine the accuracy specifications $\mathcal{E}_j(\iota\omega)$, namely $W_j(\iota\omega)$ and $V_j(\iota\omega)$, for the external-area ROMs $\hat G_j$, $j \in \{1, \dots, k\}$, many established MOR methods can be used to reduce the external-area subsystems until those specifications are satisfied.
In cases where the subsystem model is known, balanced truncation (BT) can be implemented iteratively by increasing the order of the ROM until the resulting ROM satisfies $\|W_j(\iota\omega) (\hat{G}_j(\iota\omega) - G_j(\iota\omega)) V_j(\iota\omega)\|_2 < 1$ for all $\omega \in \Omega$.
Variations, such as frequency-weighted balanced truncation (FWBT) can also be applied in a similar way.

However, for cases where only data rather than the full model of the subsystems is available, a data-driven method is required.
For example, the equipment provider of the IBRs in a sub-grid may not disclose its detailed mathematical model for confidentiality reasons but usually will provide a black-box model whose frequency scans are accessible.
In this paper, we show how our method can be used in this case.
To this end, we equip the Loewner framework~\cite[Chapter~$8$]{benner2017model} with the greedy algorithm by iteratively increasing the complexity of the ROM until it satisfies the accuracy specification $\|W_j(\iota\omega) (\hat{G}_j(\iota\omega) - G_j(\iota\omega)) V_j(\iota\omega)\|_2 < 1$ for all $\omega \in \Omega$.
We summarise this method in Algorithm~\ref{alg:Loewner} and provide the MATLAB code in~\cite{zhang2026github}.
Algorithm~\ref{alg:Loewner} is reminiscent of the AAA algorithm~\cite{nakatsukasa2018aaa} and can be interpreted as a greedy heuristic that reduces the dominant peaks of the frequency-weighted reduction errors, rapidly resulting in a reduced-order subsystem model that satisfies the accuracy specification.

\begin{algorithm}[t]
\caption{Loewner framework with greedy algorithm}\label{alg:Loewner}
\begin{algorithmic}[1]
\Require The frequency response data $G_j(\iota \omega)$ and reduction accuracy specifications $W_j(\iota\omega)$ and $V_j(\iota\omega)$, for $\omega \in \Omega$.
\State Select two frequencies related to $G_j$'s peak magnitudes.
\State Add these frequencies (in conjugate pair) and the associated frequency response data into the Loewner data set. 
\State Construct the state-space $\hat G_j$ following~\cite[Sect.~$8.3.1$]{benner2017model}, while reflecting any of $\hat G_j$'s unstable poles without changing the associated residues.
\While{$\hat G_j$ does not satisfy its accuracy specification~\eqref{eq:errSpecSubSys}}
\State Find $\omega$ where $\|W_j(\iota\omega) (\hat{G}_j(\iota\omega) - G_j(\iota\omega)) V_j(\iota\omega)\|_2$ is the largest across $\Omega$. Repeat $2$--$3$.
\EndWhile
\Ensure ROM $\hat G_j$ satisfying the accuracy specification.
\end{algorithmic}
\end{algorithm}

\subsection{Overview of Modular Model Order Reduction Framework}\label{sec:subOverviewModularMOR}
The overall modular MOR framework is summarised as follows.
\begin{enumerate}[(1)]
    \item Represent the overall interconnected power grid $\mathcal{G}$ as the LFT in~\eqref{eq:Gc}, see also Fig.~\ref{fig:FOMLFT}, after constructing the external-area subsystem collection $G_b$ following~\eqref{eq:Gb} and the dynamic interface $K$ following~\eqref{eq:interfaceK}-\eqref{eq:determineInterfaceK}.
    \item Given the user-defined accuracy specification $\mathcal{E}_c(\iota\omega)$ for the overall reduced-order interconnection, solve~\eqref{eq:LMI4Optimisation} to determine the accuracy specifications $\mathcal{E}_j(\iota\omega)$ for each external-area ROMs.
    \item Reduce each external-area subsystem independently using one of the MOR methods described in Section~\ref{sec:subDecentralisedMOR4Subsystems} such that its accuracy specification is satisfied.
    \item Construct the reduced-order interconnected power grid $\hat{\mathcal{G}}$ as in~\eqref{eq:hatGc}, see also Fig.~\ref{fig:ROMLFT}, by reassembling the reduced-order external area $\hat G_b$ with the preserved dynamic interface $K$ (including the unreduced study area).
\end{enumerate}
The resulting $\hat{\mathcal{G}}$ satisfies the prescribed accuracy specification and preserves (internal) stability provided that all the external-area ROMs satisfy their derived accuracy specifications.

In summary, we reduce the overall interconnected power grid in a modular way by deriving and enforcing subsystem accuracy specifications based on the prescribed accuracy specification on the overall interconnection, thus addressing Problems~\ref{prob:I} and~\ref{prob:III}.

\section{Application to IEEE 118-Bus Benchmark Model}\label{sec:case-study}

In this section, we present a comprehensive case study using a modified IEEE $118$-bus benchmark to demonstrate the efficacy and performance of the proposed modular MOR framework.

\subsection{Power Grid Model Reformulation}\label{sec:subPowerGridReformulation}

\begin{figure}[!t]
\centering
\includegraphics[width=\columnwidth]{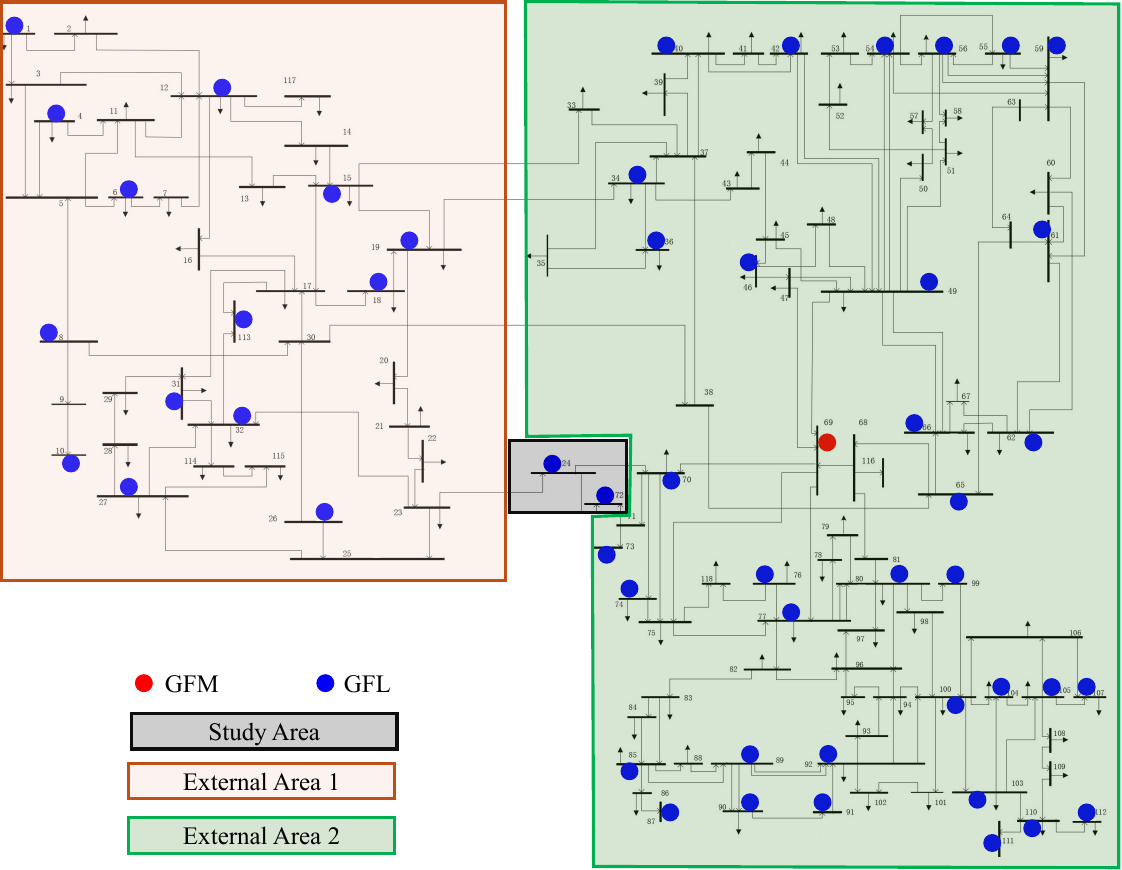}
\caption{A modified IEEE $118$-bus benchmark partitioned into a study area (grey) and an external area consisting of two sub-grids (red and green)~\cite{Al-Roomi2015}.\label{fig:118_topology}}
\end{figure}

Based on the modified IEEE $118$-bus benchmark\footnote{The system parameters are uploaded to~\cite{zhang2026github}}. with the area partitions outlined in Fig.~\ref{fig:118_topology}, all the study area and external-area subsystems are linearised at a preset operating point.
The system configuration is summarised in Table~\ref{tab:subsystemInfo}.
Note that the $m$ inputs and $p$ outputs of the subsystems $G_0$, $G_1$, and $G_2$ correspond to the voltages and currents (in the $dq$-frame) of the transmission lines at the boundary between every two of these subsystems.
The external input $u_c$ and the external output $y_c$ of $\mathcal{G}$ are (arbitrarily) selected as the active power reference $\bar P$ and the actual active power output $P$ of GFL~$24$, respectively.
The study area $G_0$ is of much lower order compared to the external areas $G_1$ and $G_2$. This aligns with the context of MOR for power systems, where a small local study area is investigated and its detailed model is retained~\cite{chow2013power}. In contrast, a large-scale external area is linearised and reduced.

\begin{table}[!t]
    \caption{System Configuration}
    \label{tab:subsystemInfo}
    \centering
    \begin{tabular}{c c c c c c}
    \toprule
    System & Order & $m$ &  $p$ & Reduced order & Reduction ratio \\
    \midrule
    $G_0$ & $34$ & $7$ & $7$ & {$34$ (unreduced)} & -- \\
    $G_1$ & $367$ & $8$ & $8$ & {$61$} & $83.38\%$ \\
    $G_2$ & $904$ & $10$ & $10$ & {$48$} & $94.69\%$ \\
    \midrule
    $\mathcal{G}$ & $1305$ & $1$ & $1$ & {$143$} & $89.04\%$ \\
    \bottomrule
    \end{tabular}
\end{table}

To apply the modular MOR framework, we first reformulate the benchmark model into the upper LFT representation as in~\eqref{eq:Gc} (see Step~1 of Section~\ref{sec:subOverviewModularMOR}).
To this end, the decoupled external-area model $G_b = \diag(G_1, G_2)$ is collected and the dynamic interface $K$ is determined by~\eqref{eq:interfaceK}-\eqref{eq:determineInterfaceK} with {$P_B \in \RR^{7 \times 19}$, 
$P_C \in \RR^{19 \times 7}$, 
and $D_K \in \RR^{19 \times 19}$, all constructed following the explanation below~\eqref{eq:determineInterfaceK}.}


\subsection{Modular Model Order Reduction Implementation}\label{sec:subModularMORSimulation}

Given a user-defined reduction accuracy specification $\mathcal{E}_c$ for the overall interconnected power grid model, we show how the accuracy specifications, namely $\mathcal{E}_1$ and $\mathcal{E}_2$, are determined for each external-area subsystem, and how the MOR is performed at a subsystem level such that the obtained reduction accuracy specifications at the subsystem level are met.
The framework is implemented in the following steps.

\begin{figure}[!t]
\centering
\includegraphics[width=\columnwidth]{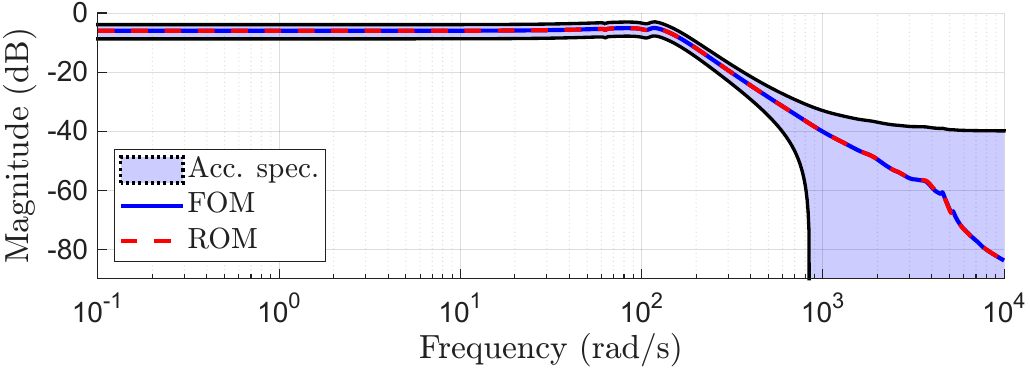}
\caption{The magnitude response of the overall interconnected power grid model $\mathcal{G}$ (solid/blue line) and its ROM $\hat{\mathcal{G}}$ (dashed/red line), and the user-defined reduction accuracy specification $|\mathcal{G}(\iota\omega)| \pm W_c^{-1}(\iota\omega)$ (shaded purple area). \label{fig:modularMORGc}}
\end{figure}

\textbf{(1) Design frequency-dependent reduction accuracy specification for the overall interconnected power grid:} 
Consider a frequency region $\Omega \subset R$, containing a fine grid of $1000$ logarithmically-spaced points in the interval $[10^{-1}, 10^4]$~rad/s.
For every frequency point $\omega \in \Omega$, a frequency-dependent reduction accuracy specification $\mathcal{E}_c(\iota\omega)$ is defined for the interconnected power grid according to~\eqref{eq:errSpecInterconnectedSys}.
In this example, we select $V_c(\iota\omega) = 1$ as constant and $W_c(\iota\omega)$ such that
\begin{equation*}
    W_c^{-1}(\iota\omega) = |0.25 \times \mathcal{G}(\iota\omega)| + 0.01.
\end{equation*}
Here, $W_c^{-1}$ is a fraction of the magnitude response of the overall interconnected power grid $\mathcal{G}$, with a small offset that allows a higher relative reduction error at higher frequencies, which is an acceptable design of the accuracy specification for power systems.
The shaded purple area in Fig.~\ref{fig:modularMORGc} depicts the corresponding magnitude envelope of the designed accuracy specification for the interconnected power grid.
One can see that the practical implication of the accuracy specification is an admissible shaded area around the nominal response of the power grid, where the ROM response should lie.

\textbf{(2) Solve for the largest admissible sets of reduction accuracy specifications for the external-area subsystems:}
Provided the predefined accuracy specification $W_c(\iota\omega)$ and $V_c(\iota\omega)$, \eqref{eq:LMI4Optimisation} is solved for all $\omega \in \Omega$, which yields $W_j(\iota\omega)$ and $V_j(\iota\omega)$ for $j = \{1, 2\}$.
Practically, this calculation is carried out by solving a bilinear optimisation problem using, \eg, the D-K iteration~\cite{zhou1998essentials}.
Then, the reduction accuracy specifications $\mathcal{E}_1(\iota\omega)$ and $\mathcal{E}_2(\iota\omega)$ for the external-area subsystems $G_1$ and $G_2$, respectively, are constructed as {in}~\eqref{eq:errSpecSubSys}.
Thanks to the specific structure of $W_j(\iota\omega)$ and $V_j(\iota\omega)$, namely being real-valued diagonal matrices with positive entries as assumed in Theorem~\ref{thm:VWMainLoopThm}, each pair of the diagonal entries in the obtained $W_j(\iota\omega)$ and $V_j(\iota\omega)$ describes the reduction accuracy specification of the associated individual input-output pair.

\begin{figure}[!t]
\centering
\includegraphics[width=\columnwidth]{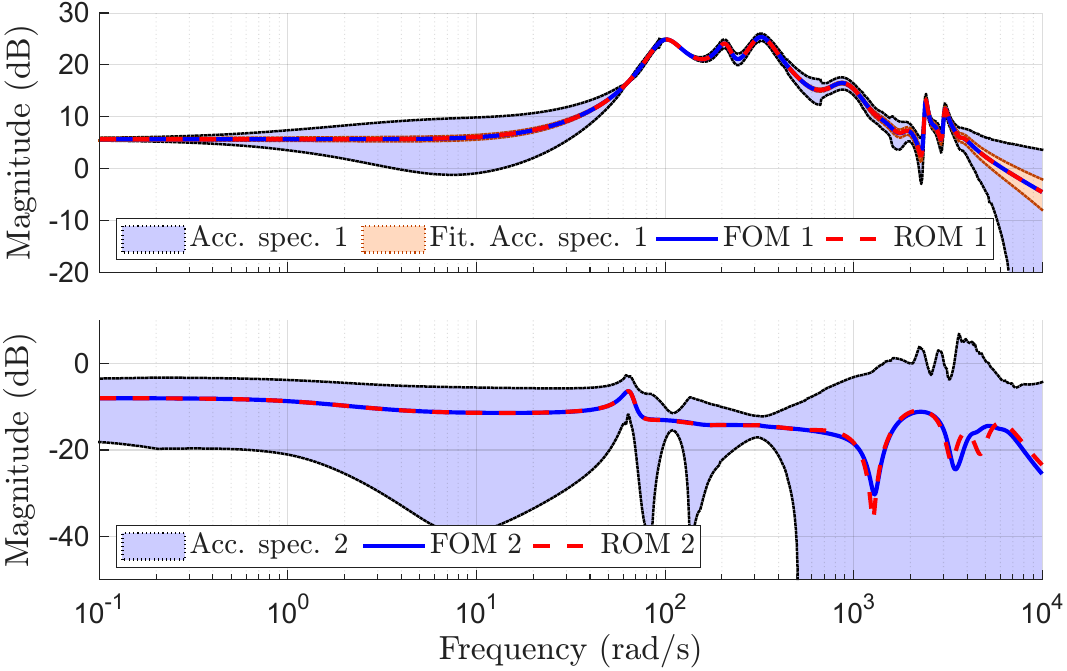}
\caption{Top: the $(7,7)$ components of the magnitude responses of $G_1$ (solid/blue line) and $\hat G_1$ (dashed/red line). The shaded purple and orange areas illustrate the original and fitted accuracy specifications for the selected input-output channel, respectively. Bottom: the $(2,1)$ components of the magnitude responses of $G_2$ (solid/blue line) and $\hat G_2$ (dashed/red line). \label{fig:modularMORG2}}
\end{figure}

\textbf{(3) Apply MOR to the external-area subsystems in a modular manner:}
As already mentioned, the proposed framework enjoys the flexibility that different MOR methods can be applied to different external-area subsystems in different contexts, as long as the obtained ROMs satisfy their associated reduction accuracy specifications.
This property enables the use of classical but intrusive methods when the subsystem model is fully known, such as balanced truncation, or the use of data-driven methods when only data are available for the subsystem model.
To illustrate this flexibility, we employ the frequency-weighted balanced truncation method~\cite{enns1984model} to reduce the subsystem $G_1$ (of order $367$) and the proposed greedy Loewner framework, \ie, Algorithm~\ref{alg:Loewner}, to reduce the subsystem $G_2$ (of order $904$), thus assuming that only the state-space model of $G_1$ is known, while for $G_2$ only its frequency response data on $\Omega$ are available.
Note that reducing $G_1$ requires the input and output weighting systems (in state-space form) for frequency-weighted balanced truncation, which can be naturally selected as the accuracy specifications $V_1$ and $W_1$ of our method.
In this case study, the weighting functions $V_1$ and $W_1$ are estimated using a stable bi-proper minimum-phase system of order~$3$ which is fitted to the sampled frequency-dependent data points $V_1(\iota\omega)$ and $W_1(\iota\omega)$, respectively, see~\cite[Chapter~$6.5$]{boyd2004convex} for details.
Every diagonal entry of the fitted weighting functions $V_1$ and $W_1$ is forced to be bounded from below by the values $V_1(\iota\omega)$ and $W_1(\iota\omega)$ for all $\omega \in \Omega$.
This introduces conservativeness, as shown by the orange shaded area in the top plot of Fig.~\ref{fig:modularMORG2}, but guarantees $G_1$'s satisfaction of its accuracy specification on $\Omega$.

By following these three steps, the subsystem $G_1$ is reduced from order $367$ to $61$ ($83.38\%$ reduction), and the subsystem $G_2$ is reduced from order $904$ to $48$ ($94.69\%$ reduction).
Consequently, the overall interconnected power grid model is reduced from $1305$ to $143$ ($89.04\%$ reduction).
These details are collected in Table~\ref{tab:subsystemInfo}.
Finally, we mention that to further illustrate the flexibility of the method, we have tested additional MOR techniques. These additional comparisons can be found in Section~\ref{sec:subComparisonDiffMORTechniques}.

The modular MOR results are presented in Figs.~\ref{fig:modularMORGc}-\ref{fig:modularMORG2}.
It is noted from Fig.~\ref{fig:modularMORGc} that the (overall) user-defined reduction accuracy specification $\|W_c(\iota\omega) E_c(\iota\omega) V_c(\iota\omega)\|_2 \leq 1$, \ie, $\hat{\mathcal{G}}(\iota\omega) \in \mathcal{E}_c(\iota\omega)$, for $\omega \in \Omega$ is achieved, which can be inferred from the magnitude response of the ROM lying in the admissible shaded area.
Similar results of the external-area subsystems $G_1$ and $G_2$ are illustrated in Fig.~\ref{fig:modularMORG2}.
We mention here that since $G_1$ and $G_2$ are multi-input multi-output, there are $164$ frequency response plots that could be reported. For reasons of space, we only show the magnitude responses of $G_1$ ($\hat G_1$) from the $7$-th input to the $7$-th output and of $G_2$ ($\hat G_2$) from the $1$-st input to the $2$-nd output.
Similar results can be concluded for all the other $162$ input-output pairs (which we do not show for space reasons). 
Consequently, it shows that the external-area ROMs $\hat G_j$ satisfy the reduction accuracy specifications $\|W_j(\iota\omega) E_j(\iota\omega) V_j(\iota\omega)\|_2 < 1$, \ie, $\hat G_j(\iota\omega) \in \mathcal{E}_j(\iota\omega)$, for $\omega \in \Omega$ and $j \in \{1, 2\}$.

In summary, for this benchmark model, we successfully determined the reduction accuracy specifications at the subsystem level that (1) guarantee that the user-defined requirements on the reduction accuracy of the overall reduced-order interconnected power grid (unreduced study area with reduced external area) are satisfied and (2) enable a significant reduction of the subsystem models, thus reducing the order of the overall interconnected power grid from $1304$ to $143$ within the predefined accuracy specification.
Moreover, the reduced-order interconnected power grid preserves its internal stability as illustrated in the next section.

\subsection{Stability Preservation and Dominant Behaviours}

In this section, we numerically validate the theoretical guarantee of (internal) stability for the reduced-order interconnected power grid $\hat{\mathcal{G}}$ and the preservation of dominant modes (when the user-defined accuracy specification is sufficiently tight) through plots of pole locations.
In addition, to demonstrate the advantages of these theoretical guarantees, we use vector fitting, a widely adopted identification technique in power systems, to obtain surrogates of the external areas $G_1$ and $G_2$, and compare the results with our method.
Vector fitting is implemented using the MATLAB function \texttt{rationalfit}, in which $600$ frequency response data (logarithmically spaced on $[10^{-1}, 10^2]$~rad/s) are used with an error tolerance of $-20$ dB.

\begin{figure}[!t]
\centering
\includegraphics[width=\columnwidth]{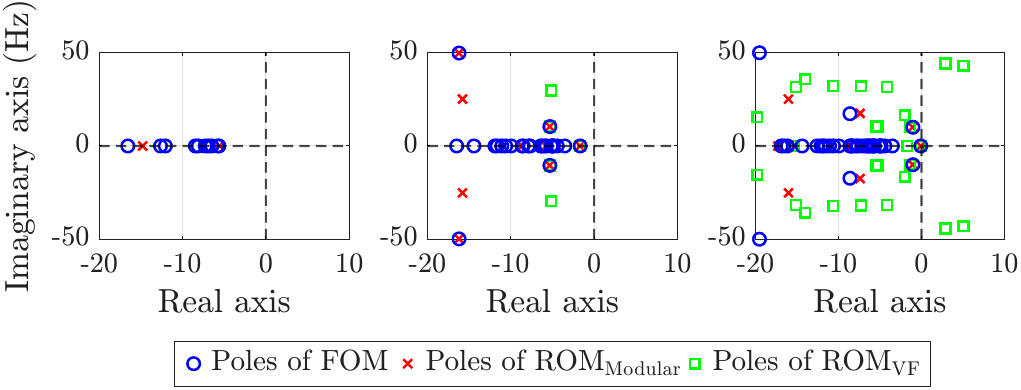}
\caption{Pole locations of the external area~$1$ (left), the external area~$2$ (middle), and the overall interconnected power grid (right). All poles of the FOMs (blue discs), the ROMs resulting from our method (red crosses), and the reduced-order surrogates resulting from vector fitting (green squares) lie in the open left half-plane, except for those of the overall interconnected surrogate arising from vector fitting.\label{fig:poleGc}}
\end{figure}

The left and middle plots in Fig.~\ref{fig:poleGc} present the pole locations of the two external areas in terms of their FOMs (blue discs), ROMs resulting from our proposed methods (red crosses), and reduced-order surrogates resulting from vector fitting (green squares).
Note that because the plots are zoomed in to show the detail around $0$, several blue discs, red crosses, and green squares are not shown.
The figure shows that our method and vector fitting both preserve the stability of the external-area subsystems and the dominant modes of the external area~$2$.
Our method also precisely captures the dominant modes of the external area~$1$.
The right plot compares the pole locations of the linear FOM $\mathcal{G}$ (blue discs), the linear ROM $\hat{\mathcal{G}}$ (red crosses), and the interconnected surrogate model (green squares).
We note that the interconnected surrogate model exhibits unstable poles even though the individual external-area surrogates \textit{per se} are stable.
However, the reduced-order interconnected power grid model obtained by our proposed framework preserves stability after interconnection, and its eigenvalues closely overlap with those of the original linearised power grid model near the imaginary axis.
In particular, the dominant pole (the pole with the largest real part) of $\hat{\mathcal{G}}$ is $-0.083$, which is close to $-0.087$, the dominant pole of $\mathcal{G}$.
It can be concluded that the obtained ROM $\hat{\mathcal{G}}$ not only preserves stability but also captures the dominant behaviour of the original FOM $\mathcal{G}$.
This observation reflects the advantage of the stability guarantee provided by our proposed framework and the risk of an unstable overall interconnection when using standalone state-of-the-art model reduction or system identification methods.

\subsection{Time-Domain Validation: Oscillatory Behaviour under Disturbances}

In this section, we investigate the oscillatory behaviour of the interconnected power grid under small local disturbances.
Two cases are demonstrated to validate the reduced-order interconnected power grid model obtained from the proposed modular MOR framework.

\textbf{Case I (Active power reference disturbance):} We denote the original nonlinear and linearised model of the interconnected power grid as $\text{FOM}_\text{NL-NL}$ and $\text{FOM}_\text{Lin-Lin}$, respectively.
In this case, we reconnect the (linear) external-area ROMs to the linearised unreduced study area and denote this reconnection as $\text{ROM}_\text{Lin-Lin}$.
A disturbance is injected at GFL~$24$, increasing its active power reference by $3\%$ at $0.1$~s.

\begin{figure}[!t]
\centering
\includegraphics[width=\columnwidth]{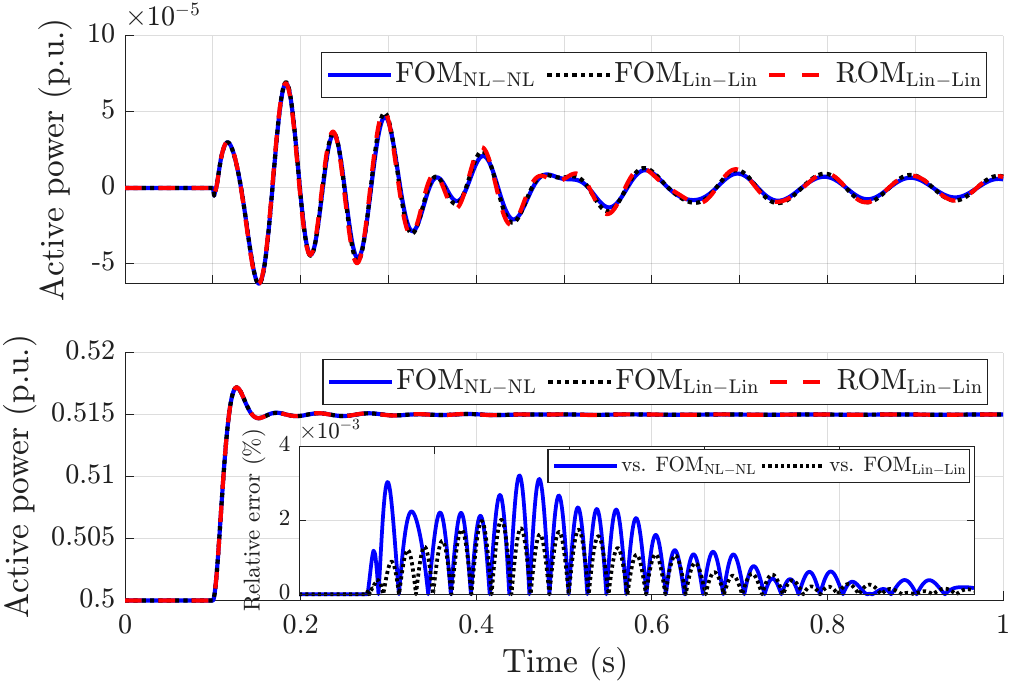}
\caption{Time histories of the active power output from GFL~$72$ (top) and GFL~$24$ (bottom), generated by $\text{FOM}_\text{NL-NL}$ (solid/blue line), $\text{FOM}_\text{Lin-Lin}$ (dotted/black line), and $\text{ROM}_\text{Lin-Lin}$ (dashed/red line). Inset: corresponding relative errors of GFL~$24$.\label{fig:timeHistoryPref2PLin}}
\end{figure}

Fig.~\ref{fig:timeHistoryPref2PLin} presents, in the top and bottom plots, the active power output (per unit) of GFL~$72$ and GFL~$24$, respectively, generated by $\text{FOM}_\text{NL-NL}$ (solid/blue line), $\text{FOM}_\text{Lin-Lin}$ (dotted/black line), and $\text{ROM}_\text{Lin-Lin}$ (dashed/red line), with a base power of $100$~MW.
The inset plot depicts the corresponding relative percentage error between the responses of $\text{ROM}_\text{Lin-Lin}$ ($P_\text{ROM}$) and both FOMs ($P_\text{FOM}$), \ie, $\frac{|P_\text{ROM} - P_\text{FOM}|}{|P_\text{FOM}|}\times 100\%$ for GFL~$24$.
The top plot shows that the output response of GFL~$72$ is well replicated by its $\text{ROM}_\text{Lin-Lin}$.
Moreover, the $10$~Hz sub-synchronous oscillatory mode\footnote{This mode is associated with the conjugate pair that is the closest to the imaginary axis, denoted by blue circles in the right plot of Fig.~\ref{fig:poleGc}.} of the overall interconnected power grid is also retained by $\text{ROM}_\text{Lin-Lin}$.
From the bottom plot, we also notice that throughout the simulation time window, $\text{ROM}_\text{Lin-Lin}$ maintains a relative error of less than $0.004\%$ compared to both {$\text{FOM}_\text{NL-NL}$} and $\text{FOM}_\text{Lin-Lin}$, with the error decaying below $0.001\%$ within $0.7$~s.
Hence, these results validate that the obtained $\text{ROM}_\text{Lin-Lin}$ accurately captures the output behaviour of GFL~$24$ for both the full-order interconnected power grid models considered in this case.

\textbf{Case II (Self-clearing inverter blocking):} In this case, we reconnect both the linearised external-area subsystems and their ROMs to the detailed nonlinear (unreduced) study area and denote them as $\text{FOM}_\text{Lin-NL}$ and $\text{ROM}_\text{Lin-NL}$, respectively.
We investigate the oscillatory behaviour of the study area when $10$~ms self-clearing inverter blocking occurs at GFL~$24$, \ie, the inverter current is blocked at $0.1$~s and restored at $0.11$~s.

\begin{figure}[!t]
\centering
\includegraphics[width=\columnwidth]{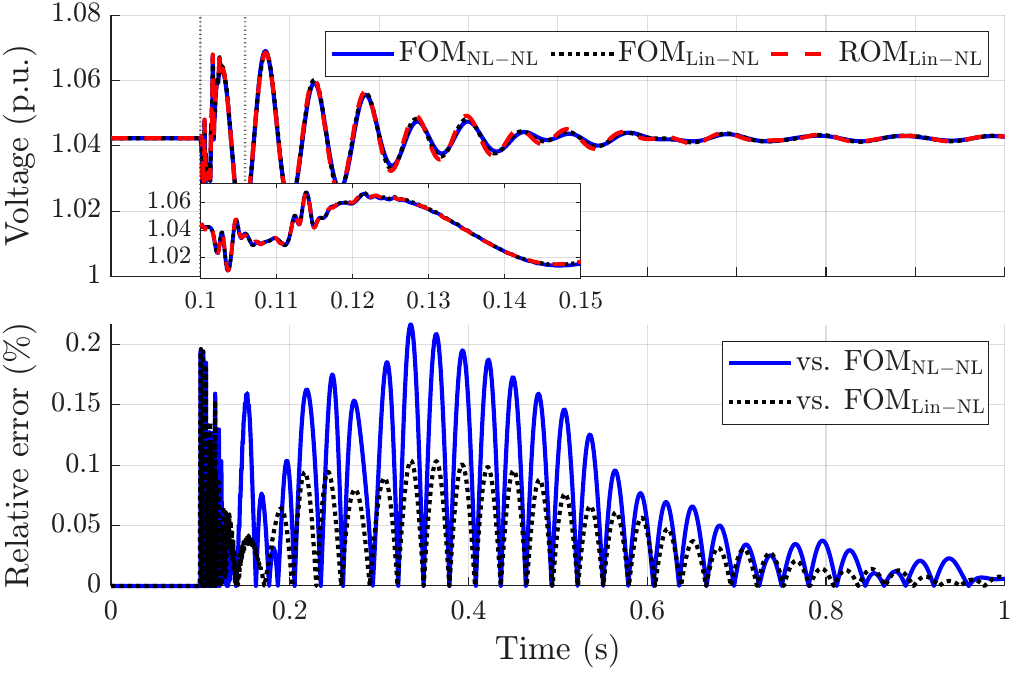}
\caption{Top: time histories of the voltage at Bus~$72$ generated by $\text{FOM}_\text{NL-NL}$ (solid/blue line), $\text{FOM}_\text{Lin-NL}$ (dotted/black line), and $\text{ROM}_\text{Lin-NL}$ (dashed/red line). Bottom: corresponding relative errors.\label{fig:timeHistoryPref2P}}
\end{figure}

The voltages of Bus~$72$, the point-of-connection between the external areas $1$ and $2$, generated by $\text{FOM}_\text{NL-NL}$ (solid/blue line), $\text{FOM}_\text{Lin-NL}$ (dotted/black line), and $\text{ROM}_\text{Lin-NL}$ (dashed/red line) are collected in the top plot of Fig.~\ref{fig:timeHistoryPref2P}, and the corresponding relative errors are depicted in the bottom plot.
It is noted that the bus voltage drops by $2.91\%$ and recovers within $0.5$~s after the fault is cleared.
The bus voltages of $\text{ROM}_\text{Lin-NL}$ and both FOMs are almost indistinguishable when GFL~$24$ encounters the blocking.
The bottom plot in Fig.~\ref{fig:timeHistoryPref2P} shows the respective relative errors in percentage.
The largest error occurs in the transient period after the bus voltage drops (\ie, $t = 0.34$~s).
This peak error is small (\ie, $0.21\%$) and decreases to less than $0.1\%$ within $0.3$~s.
We conclude that when self-clearing inverter blocking occurs at a GFL in the power grid, the point-of-connection voltage is precisely reproduced by the resulting $\text{ROM}_\text{Lin-NL}$.
Thus, the oscillatory behaviour is fully preserved through the proposed modular MOR framework.
Although this framework was operated based on the linearised study area, the simulations confirm that the resulting $\text{ROM}_\text{Lin-NL}$ can accurately replicate the output trajectory of both $\text{FOM}_\text{NL-NL}$ and $\text{FOM}_\text{Lin-NL}$ when the nonlinear study area is in the loop.

\subsection{Comparison of Computation Time}

To evaluate the computational efficiency of the resulting reduced-order power grid model, we compare the computational times for the different interconnections described in Cases I and II of the previous section.
In particular, we consider here the original nonlinear interconnected power grid and the interconnections between the linearised external-area FOM, the (linear) external-area ROM, linearised (unreduced) study area, and nonlinear (unreduced) study area.

Table~\ref{tab:FOMROMCompare} compares the different interconnections in terms of order and the elapsed time for a simulation of $1$~s.
All time-domain simulations were run in MATLAB/Simulink R2025b on the same personal laptop (Windows 11 Enterprise, Intel Core i7-11370H, and 16GB RAM).
Interconnections comprising purely linear components were simulated with a fixed time step of $5\times10^{-5}$~s, while interconnections comprising nonlinear subsystems were simulated using a variable-step ode45 solver with a maximum step of $10^{-4}$~s and an absolute/relative tolerance of $10^{-12}$.
Every experiment was run $10$ times, and the average elapsed time is reported.

\begin{table}[!t]
    \caption{Computational Comparison}
    \label{tab:FOMROMCompare}
    \centering
    \begin{tabular}{c c c c}
    \toprule
    Model 
    & Order
    & \multicolumn{2}{c}{Elapsed time ($s$)} \\
    \midrule
    NL. Power Grid & $1305$ & \multicolumn{2}{c}{$112.49$} \\
    \cmidrule(lr){3-4}
     & &  Lin. study area & NL. study area \\
    \cmidrule(lr){3-4}
    Lin. Ext. FOM & $1305$ & {$5.78$} & {$13.20$}  \\
    Lin. Ext. ROM & {$143$} & {$0.13$} & {$8.63$}  \\
    \bottomrule
    \end{tabular}
\end{table}


The table shows that for a simulation of $1$~s, the interconnection of the (linear) external-area ROM and the linear study area requires only $0.13$~s, which is roughly $44$ times more computationally efficient than the interconnection of the linearised external-area FOM and the linear study area, and almost $865$ times more computationally efficient than the original nonlinear interconnected power grid model.
When reconnected to the nonlinear study area, the interconnection with (linear) external-area ROM requires $8.63$~s for a simulation of $1$~s, a reduction of roughly $35\%$ relative to the interconnection with the linearised external-area FOM and roughly $93\%$ relative to the original nonlinear interconnected power grid model, even though the computation time of the nonlinear study area is dominant in this configuration.

In conclusion, reducing the order of the overall interconnected power grid from $1305$ to $143$ substantially reduces the simulation time, resulting in a computationally efficient ROM of the overall interconnected power grid, while providing rigorous accuracy and stability guarantees.

\subsection{Comparison of Different Model Order Reduction Techniques}\label{sec:subComparisonDiffMORTechniques}

We evaluate the performance of different combinations of subsystem MOR techniques and show how different selections affect the orders and system errors of the obtained ROMs.

We observe from Table~\ref{tab:MORCompare} that our proposed greedy Loewner framework achieves relatively small frequency-weighted system errors at the cost of reducing the subsystem order to a value between what balanced truncation (BT) and frequency-weighted balanced truncation (FWBT) achieve.
Nevertheless, it outperforms the other methods in extraction time, \ie, the time required to compute the ROM, especially for $G_2$ whose original FOM has large order.
This outperformance arises from the data-driven nature of the approach, while BT and FWBT need to compute Gramians, which is expensive for systems of very high order.
Furthermore, it achieves the smallest weighted $\mathcal{H}_\infty$ error at the subsystem level, which is consistent with the nature of its embedded greedy algorithm.

\begin{table*}[!t]
    \caption{Comparison of Different Model Order Reduction Techniques}
    \label{tab:MORCompare}
    \centering
    \begin{tabular}{c c c c c c c c c c}
    \toprule
    \multirow{2}{*}[-0.5ex]{Method}
    & \multicolumn{2}{c}{Order}
    & \multicolumn{2}{c}{Extraction time ($s$)}
    & \multicolumn{3}{c}{Weighted (unweighted) {$\mathcal{H}_\infty$} error}
    & \multicolumn{2}{c}{Dominant pole within $\pm 50$~Hz band} \\
    \cmidrule(lr){2-3}
    \cmidrule(lr){4-5}
    \cmidrule(lr){6-8}
    \cmidrule(lr){9-10}
    & {$\hat G_1$} & {$\hat G_2$}
    & {$\hat G_1$} & {$\hat G_2$}
    & {$\hat G_1$} & {$\hat G_2$} & {$\hat{\mathcal{G}}$}
    & $G_1 (-5.64)$ & $G_2(-1.67)$ \\
    \midrule
    BT & {$73$} & {$62$} & {$31.10$} & {$294.92$} & {$0.70\, (0.038)$} & {$0.93\, (0.033)$} & {$0.043\, (0.0061)$} & $-6.05$ & $-1.70$ \\
    FWBT & {$61$} & {$44$} & {$32.33$} & {$256.98$} & {$0.81\, (0.093)$} & {$0.88\, (0.14)$} & {$0.021\, (0.0030)$} & {$-5.54$} & {$-1.70$} \\
    Greedy Loewner & {$82$} & {$48$} & {$22.98$} & {$11.45$} & {$0.48\, (0.20)$} & {$0.56\, (0.24)$} & {$0.074\, (0.011)$} & {$-0.94 \pm 12.35\iota$} & {$-1.67$} \\
    \bottomrule
    \end{tabular}
\end{table*}

\section{Conclusion}\label{sec:conclusion}

A modular MOR framework has been proposed in this paper.
Given a user-defined reduction accuracy specification for the overall interconnected power grid, the reduction accuracy specifications for the external-area subsystems are determined.
Each external-area subsystem is then reduced individually, with no need to share model information with others.
The proposed framework preserves the interconnection structure of the original interconnected power grid, enhancing the interpretability and physical consistency of the resulting reduced-order interconnected power grid model.
Both (frequency-weighted) balanced truncation and the newly developed greedy Loewner framework have been applied to obtain external-area ROMs that satisfy the corresponding accuracy specifications.
The obtained external-area ROMs are finally reassembled and then reconnected to the unreduced study area, resulting in a reduced-order interconnected power grid model that satisfies the predefined reduction accuracy specification, preserves stability, and requires less computation time for simulation.

\bibliographystyle{IEEEtran}
\bibliography{ref}











\newpage

 




\vfill

\end{document}